\documentclass[times,authoryear]{elsarticle}

\usepackage{jasr}
\usepackage{framed,multirow}

\usepackage{amssymb}
\usepackage{latexsym}

\usepackage[switch]{lineno}
\usepackage{amsmath}
\usepackage{url}
\usepackage{xcolor}
\definecolor{newcolor}{rgb}{.8,.349,.1}

\usepackage[citebordercolor=white]{hyperref}

\journal{Advances in Space Research}

\begin{document}

\verso{Given-name Surname \textit{etal}}

\begin{frontmatter}

\title{Statistical Equivalence of Metrics for Meteor Dynamical Association}%

\author[1,2]{Peña-Asensio \snm{Eloy}\corref{cor1}}
\ead{eloy.pena@polimi.it, eloy.pena@uab.cat}
\cortext[cor1]{Corresponding author: 
  Tel.: +39-02-2399-7157}
  
\author[3]{Sánchez-Lozano \snm{Juan Miguel}}

\affiliation[1]{organization={Department of Aerospace Science and Technology, Politecnico di Milano},
            addressline={Via La Masa 34}, 
            city={Milano},
            postcode={20156}, 
            state={Lombardia},
            country={Italy}}

\affiliation[2]{organization={Departament de Química, Universitat Autònoma de Barcelona},
            addressline={Carrer dels Til·lers}, 
            city={Bellaterra},
            postcode={08193}, 
            state={Catalonia},
            country={Spain}}

\affiliation[3]{organization={Centro Universitario de la Defensa, Universidad Politécnica de Cartagena},
                addressline={C/ Coronel López Peña S/N, Base Aérea de San Javier},
                city={Santiago de la Ribera},
                postcode={30720},
                country={Spain}}

\received{1 May 2013}
\finalform{10 May 2013}
\accepted{13 May 2013}
\availableonline{15 May 2013}
\communicated{S. Sarkar}

\begin{abstract}
Meteor showers, originating as a result of the activity of comets or the disruption of large objects, provide a unique window into the composition and dynamics of our Solar System. While modern meteor detection networks have amassed extensive data, distinguishing sporadic meteors from those belonging to specific meteor showers remains challenging. In this study, we statistically evaluate and compare four orbital similarity criteria within five-dimensional parameter space ($D_{SH}$, $D_D$, $D_H$, and $\varrho_2$) to study dynamical associations using the already classified meteors (manually by a human) in CAMS database as a benchmark. In addition, we assess various distance metrics typically used in Machine Learning with two different vectors: ORBIT, grounded in heliocentric orbital elements, and GEO, predicated on geocentric observational parameters. To estimate their degree of correlation and efficacy, the Kendall rank correlation coefficient and the Top-k accuracy are employed. The statistical equivalence of the Top-1 results is examined using the Kolmogorov-Smirnov test and the percentage of Top-1 agreement is calculated on an event-by-event basis. Additionally, we compute the optimal cut-offs for all methods for distinguishing sporadic background events. Our findings demonstrate the superior performance of the sEuclidean metric in conjunction with the GEO vector. Within the scope of D-criteria, $D_{SH}$ emerged as the preeminent metric, closely followed by $\varrho_2$. The Bray-Curtis metric displayed an advantage compared to the other distance metrics when paired with the ORBIT vector for Top-k accuracy, however, the Cityblock metric is more effective when considering the sporadic background. $\varrho_2$ stands out as the most equivalence to the distance metrics when utilizing the GEO vector and the most compatible with GEO and ORBIT simultaneously, whereas $D_D$ aligns more closely when using the ORBIT vector. The stark contrast in $D_D$'s behavior compared to other D-criteria highlights potential inequivalence. Our results suggest that geocentric features provide a more robust basis than orbital elements for meteor dynamical association. Most distance metrics associated with the GEO vector surpass the D-criteria when differentiating the meteoroid background. Accuracy displayed a dependence on solar longitude with a pronounced decrease around 180$^\circ$ matching an apparent increase in the meteoroid background activity, tentatively associated with the transition from the Perseids to the Orionids. Considering lately identified meteor showers, $\sim$27\% of meteors in CAMS would have different associations. This work unveils that Machine Learning distance metrics can rival or even exceed the performance of tailored orbital similarity criteria for meteor dynamical association.

\end{abstract}

\begin{keyword}
\KWD Meteor\sep Meteoroids\sep Meteor showers\sep Meteoroid streams\sep Statistics
\end{keyword}

\end{frontmatter}


\section{Introduction}

Within the expanse of our planetary system, remnants from its formation provide glimpses into the early stages of our cosmic neighborhood \citep{Bottke2002, Walker2006book}. Among these remnants, comets emerge as witnesses to the dramatic events that shaped our nearby environment. These celestial bodies can undergo processes of disruption due to various factors such as volatile sublimation when approaching the Sun, tidal forces, or impacts with other bodies. According to the theory of formation and evolution of small bodies of the Solar System \citep{Whipple1951, Bredikhin1954, Plavec1954, Hughes1986, Babadzhanov1992}, meteoroid streams are formed mainly as a result of the activity of comets or the ejection of meteoroids from cometary nuclei with various initial velocities \citep{Chapman2010, Toth2011, Gritsevich2012CosRe}. Meteoroids exhibit a diverse composition, including rock, metal, or a combination of both, and span a wide range of sizes, from micrometer-scale grains to larger objects up to one meter in diameter \citep{Trigo2006MNRAS, Trigo2007MNRASerra, Koschny2017JIMO...45...91K}. Despite their heterogeneous characteristics, these meteoroids share a common origin, derived from a parent body, which imparts certain similarities among them.

Additionally, though less common, asteroids can also generate meteoroid streams as a result of catastrophic impact events. Some associations have been found, such as the case of the potentially hazardous asteroid (3200) Phaethon (1983TB), whose origin could be the nucleus of an extinct comet \citep{LIN2020}, and the Geminids meteor shower. Multiple studies have confirmed the high probability that the Geminids are dynamically associated with such asteroid \citep{Whipple1983, Gustafson1989, Williams1993}. However, as they traverse the space, the influence of planetary perturbations and non-gravitational forces gradually renders them indistinguishable from the background population \citep{Olsson1986, Bottke2000, PaulsGladman2005, Broz2006, Koschny2019}.

Eventually, the journey of meteoroids brings them into intersecting paths with the Earth's orbit, leading to captivating interactions with our planet \citep{Brown2002, Murad2002, Gritsevich2009AdSpR, Trigo2022}. As these meteoroids penetrate the Earth's atmosphere, they experience a dramatic transformation fueled by the intense heat generated through air molecule friction. The high-speed entry produces enormous amounts of heat, causing the outer layers of the meteoroids to rapidly vaporize \citep{Popova2019msmebook}. This process, known as ablation, leads to the formation of a glowing plasma sheath surrounding the meteoroid \citep{Ceplecha1998SSRv, Silber2018AdSpR}. The energy released during atmospheric aerobraking causes the visible phenomenon known as a meteor, which is called a fireball or bolide if its magnitude surpasses that of the planet Venus. When a meteoroid stream intersects the Earth's path periodically, it gives rise to the phenomenon of meteor showers \citep{Jenniskens1994, Jenniskens1998, Jenniskens2006, Vaubaillon2019, Jenniskens2023Atlas}. The meteors within them share common features, including their time of occurrence, apparent origin in the sky, known as the radiant, and their geocentric impact velocity, as well as their orbital elements in an equivalent manner. 

Thanks to the existence of meteor detection networks \citep{Jacchia1956, Ceplecha1957, Bland2004, trigo2005a, Weryk2007, Jenniskens2011Icar, Kornos2014pim3, Gritsevich2014, Colas2014old, SonotaCo2016JIMO, Gardiol2016, Devillepoix2020, Colas2020AA, Boaca2022ApJ, Borovicka2022AA}, researchers now have access to an abundance of videos and images that play a key role in improving our understanding of the distribution and characteristics of meteoroid streams. The availability of this extensive dataset has revolutionized our capacity to investigate and analyze incoming extraterrestrial material. Through processing these videos and extracting relevant information, we can determine essential parameters such as trajectory, velocity, brightness, and other physical properties of meteors \citep{Ceplecha1987, Borovicka1990BAICz, Vida2020MNRAS, Eloy2021MNRAS, Eloy2024Icar}. This valuable information significantly contributes to our knowledge about meteor showers, assisting in the identification and tracking of potentially hazardous complexes that may pose a threat to Earth in the near term \citep{Voloshchuk1996, Halliday1987, Borovi2015, Trigo2017, TrigoBlum2022MNRAS}.

In this regard, accurate classification of sporadic meteors from those belonging to the same stream is of utmost importance, as the latter provides valuable insights into the population density of future impactors \citep{Wiegert2004EMP, Porub2004EM&P...95..697P, Jopek2013MNRAS, Dumitru2017AA, Jenniskens2017PSS, Vaubaillon2019, Eloy2022AJ, Eloy2023MNRAS}. Determining the point at which a meteor shower transitions from a cohesive entity to a collection of unrelated meteoroids (sporadic background), or establishing the criteria to accurately associate meteors with a specific shower, poses a significant challenge. To tackle the issue of orbital dynamical association, multiple endeavors have been undertaken to define similarity criteria or D-criteria. These criteria aim to effectively differentiate between events that are associated with a specific meteoroid stream and those that are unrelated to other objects or swarms. Ultimately, analyzing the impact features can aid in associating meteorites with their parent bodies \citep{Carbognani2023MNRAS}.

In this study, we assess the rank correlation, efficacy, and equivalence of four five-dimensional similarity criteria designed for quantifying dynamical associations between meteor orbits, as well as various distance metrics with two different vectors (one shared with the D-criteria). The evaluation is conducted using a comprehensive meteor database and extends to exploring alternative metrics for orbit association, as well as computing the optimal thresholds for each method. The objective is to elucidate the statistical strengths, limitations, and similarities of each approach, thereby providing a robust framework for future research in meteor associations with parent bodies or meteoroid streams.

In Section \ref{DataMethods}, we detail the database utilized and the methodology applied. Section \ref{results} presents our findings, and Section \ref{conclusions} provides a summary of the key outcomes of our study.


\section{Data and Procedures} \label{DataMethods}

The methodology presented herein is designed to analyze multiple meteor dynamical association approaches by comparing five-dimensional orbital similarity criteria and various vector-based distance metrics typically used in Machine Learning. For the latter, we use as a vector (1) the same parameters utilized by the similarity criteria defined by some heliocentric orbital elements, which we termed as ORBIT, and (2) the four-dimensional vector proposed by \citet{Sugar2017MPS} and named here as GEO. It should be noted that while the term ``metrics’’ may be appropriate to describe the D-criteria to a certain extent, in this work, we use the term ``metrics’’ exclusively to refer to vector-based distance metrics, which are further explained.

This section is subdivided into different subsections. Subsection \ref{data} elaborates on the data sources utilized. Subsection \ref{orbital-similarity} presents D-criteria for comparing the orbital elements of two orbits. In Subsection \ref{vector-distance}, we introduce the two vectors that will be used along with the distance metrics. In Subsection \ref{statistical-equivalence} we explain the theoretical background used for calculating the rank correlations, comparing the performances with the Top-k accuracy method, and estimating the equivalence with the Kolmogorov-Smirnov test and Top-1 event-by-event agreement. Finally, in Subsection \ref{diff_sporadic}, we detail our strategy to determine the optimal thresholds for distinguishing between sporadic background and meteor showers. All implementations of the statistical analyses were conducted utilizing the \textit{SciPy} library \citep{2020SciPy}.

\subsection{Databases} \label{data}

CAMS, short for the Cameras for All-Sky Meteor Surveillance project \citep{Jenniskens2011Icar}, is an international initiative sponsored by NASA and managed by the Carl Sagan Center within the SETI Institute, located in California, USA. Its primary objective is to monitor and map meteor activity through nighttime optical video surveillance, employing triangulation techniques. It annually records an average of half a million meteor orbits, although the publication of this data stopped in 2016. The last release was the Meteoroid Orbit Database v3.0, which includes 471,582 events registered since 2010.

While there are other automated meteor detection networks, CAMS stands out as the primary and most widely recognized repository of meteor data. Nevertheless, it was noted that its performance in accurately detecting fast meteors falls short in comparison to its detection of slower meteors \citep{Koseki2017eMetN, Koseki2022eMetN}. To address this issue, we implement a filtering mechanism to exclude lower-quality detections and to reduce spurious data, requiring a minimum convergence angle of 15 degrees between cameras, ensuring an estimated velocity error of no more than 10\% of the nominal value, not allowing hyperbolic orbits, and selecting perihelion distance compatible with impacts on the Earth.

Certainly, we rely on the classification provided by CAMS as a ground truth, which may not be infallible. However, the classification within this database did not utilize any formal dissimilarity criteria. Instead, it depended on human visual clustering within sun-centered ecliptic longitude-latitude representations, with clusters manually delineated using specific coordinates and geocentric velocity limits \citep{Jenniskens2018PSS}. Our analyses proceeds under the presumption that the CAMS classification is accurate, a premise that, regardless, serves our primary objective of assessing the equivalence between metrics and D-criteria.

For identifying meteoroid streams responsible for meteor showers, we use the V.2 list of all known showers from the IAU Meteor Data Center, updated in January 2024 \citep{Jopek2011msssconf, jopek2013meteoroids, Jopek2017PSS, Jenniskens2020PSS}. To facilitate the association of these meteor showers with entries in the CAMS database, we employ the IAU numeral code. This list includes 1484 entries, 956 corresponding to unique meteor showers. To ensure a direct comparison of association performances, we filter both CAMS and IAU meteor shower datasets to include only identical, unique meteor showers.

\subsection{Orbital Similarity Criteria}
\label{orbital-similarity}

Orbital elements such as inclination $i$, eccentricity $e$, longitude of the ascending node $\Omega$, perihelion distance $q$, and argument of the perihelion $\omega$ allow us to determine the path of any moving object following a Keplerian trajectory in our Solar System. Likewise, it is possible to look for the connection between a meteor shower and its parent body (or any two objects) through the similarities of their orbits.

This search approach is not recent. The first attempts focused on measuring the degree of similarity between orbits were designed in the second half of the last century, they were so-called D-criteria. The first D-criteria was introduced by \citet{Southworth1963SCoA....7..261S}:

\begin{equation}
D_{SH}^{2}=\left(e_{B}-e_{A}\right)^{2}+\left(q_{B}-q_{A}\right)^{2}+\left(2 \sin \frac{I_{AB}}{2}\right)^{2}
+\left(\frac{e_{B}+e_{A}}{2}\right)^{2}\left(2 \sin \frac{\pi_{B A}}{2}\right)^{2},
	\label{eq:D_SH}
\end{equation}

where other concepts of geometry come into play such as the angles between their respective perihelion points ($\pi_{B A}$) and between the inclinations of the orbits ($I_{AB}$).

\citet{Drummond1981Icar...45..545D} not only defined the angle between the perihelion points on each orbit ($\theta_{B A}$) by adding both the ecliptic longitude ($\lambda$) and the perihelion latitude ($\beta$), but also weighted the terms $e$ and $q$ to provide a metric in which each term contributed equally to the overall sum. In this way, a new variant of the $D_{SH}$ criterion, named $D_{D}$ in honor of its creator, was developed:

\begin{equation}
D_{D}^{2}=\left(\frac{e_{B}-e_{A}}{e_{B}+e_{A}}\right)^{2}+\left(\frac{q_{B}-q_{A}}{q_{B}+q_{A}}\right)^{2}+\left(\frac{I_{AB}}{\pi}\right)^{2}
+\left(\frac{e_{B}+e_{A}}{2}\right)^{2}\left(\frac{\theta_{B A}}{\pi}\right)^{2},
	\label{eq:D_D}
\end{equation}

A decade later, \citet{Jopek1993Icar..106..603J} carried out a random perturbation model of several orbits, ignoring $i$, $\Omega$, and $\omega$, to analyze the $D_{SH}$ and $D_{D}$ criteria. He found dependency relationships of $q$ and $e$ values for the reference orbit; $q$ in the case of $D_{SH}$ and $e$ for the criterion $D_{D}$. To reduce these dependency relationships between orbital parameters, Jopek proposed a new similarity criterion, $D_{H}$, defined by: 

\begin{equation}
D_{H}^{2}=\left(e_{B}-e_{A}\right)^{2}+\left(\frac{q_{B}-q_{A}}{q_{B}+q_{A}}\right)^{2}+\left(2 \sin \frac{I_{AB}}{2}\right)^{2}
+\left(\frac{e_{B}+e_{A}}{2}\right)^{2}\left(2 \sin \frac{\pi_{B A}}{2}\right)^{2}.
	\label{eq:D_H}
\end{equation}




Note that these D-criteria cannot be categorized mathematically as metrics due to their violation of the triangle inequality \citep{Kholshevnikov2016MNRAS}. Instead, they are more appropriately defined as quasimetrics, as they adhere to a relaxed form of the triangle inequality \citep{Milanov2019CeMDA}. Contemporary functions, such as $\varrho_2$, enable the precise quantification of orbital similarity through consistent mathematical formulations:

\begin{equation}
\varrho_2^2=\left(1+e_1^2\right) p_1+\left(1+e_2^2\right) p_2-2 \sqrt{p_1 p_2}\left(\cos I+e_1 e_2 \cos P\right)
\end{equation}
with

\begin{align}
p &= a \left(1 - e^2\right), \\
\cos I &= \cos i_1 \cos i_2 + \sin i_1 \sin i_2 \cos \left( \Omega_1 - \Omega_2 \right), \\
\cos P &= \sin i_1 \sin i_2 \sin \omega_1 \sin \omega_2 + \cos \omega_1 \cos \omega_2 \cos \left( \Omega_1 - \Omega_2 \right) \nonumber \\
&\quad + \cos i_1 \cos i_2 \sin \omega_1 \sin \omega_2 \cos \left( \Omega_1 - \Omega_2 \right) \nonumber \\
&\quad + \left( \cos i_2 \cos \omega_1 \sin \omega_2 - \cos i_1 \sin \omega_1 \cos \omega_2 \right) \sin \left( \Omega_1 - \Omega_2 \right).
\end{align}

The limit values of such D-criteria, also called thresholds, cut-off levels, or upper limits, determine whether two objects may be associated. Being, for example, $A$ and $B$ a meteor and meteor shower respectively, if the distance $D(A,B)$ between $A$ and $B$ is greater than this limit value, the association must be discarded. The smaller this distance is, the greater the possibility that there is a dynamical similarity between two objects, and, therefore, the meteoroid belongs to the meteoroid stream. 

Some studies on the suitability of these criteria have already been carried out. For example, \citet{Galligan2001} explored the performance of four similarity functions in the near-ecliptic region--$D_{SH}$, $D_D$, $D_H$, and $D_N$ \citep{Valsecchietal1999}--, resulting in $D_N$ criterion being the most stable in the case of the lack of a priori information on orbital inclination regimes, while $D_{SH}$, which is based on meteor shower dispersion theoretical models, is more suitable with very different cut-off levels. However, $D_N$ has not been adopted in our approach due to its less straightforward application from the standard parameters provided in meteor databases. Likewise, \citet{Moorhead2015} analyzed such cut-off values to determine a chosen acceptable false-positive rate and distinguish which showers are significant within a set of sporadic meteors. \citet{Jenniskens2008Icar} and \citet{Rudawska2015PSS} introduced the four-dimensional metrics $D_B$ and $D_X$, respectively. However, to maintain consistency within the parameter space domain analyzed in this study, we opt not to include these criteria.

Through these values, it has been possible to associate meteor showers with parent bodies such as the 109P (1862) III Swift–Tuttle comet and the Perseid meteor shower, whose first connection data from the late 19th century when Schiaparelli calculated the orbits of the Perseids and discovered their strong similarity to that of this comet. Involved on this connection, \citep{SOKOLOVA2014} calculated the cut-off level of $D_{SH}$ resulting in $D_{SH}$$\leqslant 0.2$. Literature provides more classical examples such as the April Lyrids, whose extremely small value of the $D_D$ criterion ($D_D$=0.009) suggests that such meteors showers have indeed come from comet Thatcher \citep{ArterandWilliams1997}. Other recent examples are the case of a fireball detected in the night sky over Kyoto whose likely parent, with $D_{SH}$= 0.0079, could be the binary near-Earth asteroid (164121) 2003 YT1 \citep{Kasuga_2020}; the binary asteroid 2000 UG11 associated with Andromedids ($D_{SH}$=0.183 and $D_H$=0.176) and the asteroid (4179) Toutatis, with values of $D_{SH}$=0.180 and $D_H$=0.175, that postulate it associated with October Capricornids \citep{Dumitru2017AA}; the meteor shower June epsilon Ophiuchids, whose values in three D-criteria ($D_{S H}$=0.05, $D_D$=0.03 and $D_H$= 0.06) confirm that is likely to originate from comet 300P/Catalina \citep{Matlovic2020}; or the recently observed fall and recovery of the Traspena meteorite is posited to be linked with the potentially hazardous asteroid 1989 QF (Minos), exhibiting $\varrho_2=$0.1059 \citep{Andrade2023MNRAS}. We note the absence of a cut-off estimate works for $\varrho_2$, unlike the traditional D-criteria.

Although the cases mentioned above demonstrate the usefulness of the similarity criteria, some limitations confirm the need to investigate these metrics. For example, \citet{Galligan2001} found that, for the case of the $D_{SH}$ criterion, it is necessary to use different upper limits depending on the orbital inclination angle of the stream. In fact, \citet{SOKOLOVA2014}, intending to improve the reliability of identification of the observed objects, recommends analyzing the $D_{SH}$ threshold values independently for each meteoroid complex. Following that approach, the study of comparison of four similarity criteria carried out by \citet{Rudawaskaetal2012} confirmed the difficulty in obtaining one specific value of threshold that would fit all cases, reaching the conclusion that the ideal threshold depends on the cluster analysis method, the meteors shower, and the sample; this latter statement is also seconded by \citet{JOPEKANDBRONIK2017}. \citet{Ye2018PSS} also pointed out that the traditional D-criteria may not necessarily reflect a shared origin of two objects due to the orbital evolution influenced by planetary perturbations.



In short, these studies are clear examples of the need to analyze the effectiveness and equivalence of the different approaches to establish dynamical associations of meteors.


\subsection{Meteor Vectors and Distance Metrics}
\label{vector-distance}

In the preceding section, we discussed five-dimensional D-criteria for associating meteors with meteor showers. While these approaches are widely used, they are not without limitations. It is an active research topic for which there is no consensus on either criteria or thresholds. To search for alternatives and compare their performance, we introduce two meteor vectors --ORBIT and GEO-- to evaluate multiple Machine Learning distance metrics in meteor-shower association.

The ORBIT vector focuses simply on the same five heliocentric orbital elements that are used by the above-mentioned orbital similarity criteria, which allows for a more direct comparison of the effectiveness:

\begin{equation}
\label{eqn:ORBIT}
ORBIT=
\left[\begin{array}{c}
q \\
e \\
i/\pi \\
\frac{\sin(\omega) + 1}{4} \\
\frac{\cos(\omega) + 1}{4} \\
\frac{\sin(\Omega) + 1}{4} \\
\frac{\cos(\Omega) + 1}{4} \\
\end{array}\right].
\end{equation}

Note that the database has been filtered to minimize spurious events, ensuring the inclusion of only non-hyperbolic orbits ($0 < e \leq 1$) that intersect Earth's orbit, specifically with $0 < q \leq 1$ au. The inclination, when normalized by 180º, spans the range [0, 1]. For the circular components, $\omega$ and $\Omega$, which range from [-1, 1], we normalize them to [0, 1] and assign half the weight to each circular component. Utilizing sine and cosine functions for the circular angle $\omega$ and $\Omega$, we effectively account for the shortest circular distance between angles, ensuring that 358$^\circ $ is recognized as 4$^\circ $ away from 2$^\circ $, rather than 356$^\circ $. Consequently, all five independent parameters are normalized and weighted equally, constructing a five-dimensional space vector.

The GEO vector is based mainly on geocentric observable parameters and was proposed by \citet{Sugar2017MPS}. This six-component vector (but in four-dimensional space as it has only four independent parameters) inherently addresses the issue of longitude wrapping. It normalizes the six components to ensure that each variable contributes equally. The vector's initial two components represent the meteor's position, as the meteoroid intersects the Earth's orbit. The subsequent three components define the unit vector opposite to the meteor's velocity direction. The final component represents the magnitude of the geocentric velocity, normalized by the maximum velocity allowed for the study population:

\begin{equation}
\label{eqn:GEO}
GEO=
\left[\begin{array}{c}
\cos \left(\lambda_{\odot}\right) \\
\sin \left(\lambda_{\odot}\right) \\
\sin \left(\lambda_{\mathrm{g}}-\lambda_{\odot}\right) \cos \left(\beta_{\mathrm{g}}\right) \\
\cos \left(\lambda_{\mathrm{g}}-\lambda_{\odot}\right) \cos \left(\beta_{\mathrm{g}}\right) \\
\sin \left(\beta_{\mathrm{g}}\right) \\
v_{\mathrm{g}} / 72
\end{array}\right].
\end{equation}

In this vector, $v_{\mathrm{g}}$ represents the geocentric velocity in kilometers per second, $\lambda_{\odot}$ is the solar longitude, $\beta_{\mathrm{g}}$ is the geocentric ecliptic latitude of the radiant, and $\lambda_{\mathrm{g}} - \lambda_{\odot}$ being the Sun-centered ecliptic longitude of the radiant. All components span the range [-1, 1], except for the element related to velocity, which varies between [0, 1]. Given that velocity measurements are subject to the greatest degree of error, the authors allowed a reduced weight for the velocity.

Although the D-criteria are theoretically five-dimensional, the orbits of the meteors are constrained by having impacted the Earth, virtually reducing the dimensionality by one. Consequently, this dimensionality reduction enables a comparison between the performances of the GEO and ORBIT vectors.

In the quest to develop a robust methodology for associating meteors with their parent meteor showers, we explore various distance metrics typically used in Machine Learning that can quantify the similarity between the previously defined vectors. In Table \ref{table:distance_metrics}, we introduce the distance metrics that are employed in this study.

\begin{table}[h]
\centering
\caption{Summary of distance metrics.}
\label{table:distance_metrics}
\begin{tabular}{ccc}
\hline
Metric Name & Formula & Brief Explanation \\
\hline
Euclidean & \(\sqrt{\sum_{i}(u_i - v_i)^2}\) & Square root of sum of squared differences \\
sEuclidean & \(\sqrt{\sum_{i}\frac{(u_i - v_i)^2} {{V_i}^2}}\) & Normalized Euclidean by variance \(V_i\) \\
Cityblock (Manhattan) & \(\sum_{i} \left | u_i - v_i \right |\) & Sum of absolute differences \\
Cosine Similarity & \(1 - \frac{u \cdot v}{\Vert u \Vert \cdot \Vert v \Vert }\) & Angle between vectors \\
Canberra & \(\sum_{i}\frac{\left | u_i - v_i \right |}{\left | u_i \right | + \left | v_i \right |}\) & Weighted Cityblock distance \\
Bray-Curtis & \(\frac{\sum_{i} \left | u_i - v_i \right |}{\sum_{i} \left | u_i + v_i \right |}\) & Normalized weighted Cityblock \\
Chebyshev & \(\max_i \left | u_i - v_i \right |\) & Maximum absolute difference \\
\hline
\end{tabular}
\end{table}

\subsection{Statistical Analysis}
\label{statistical-equivalence}

\subsubsection{Rank Correlations}

We select the Kendall rank correlation coefficient ($\tau$) to measure the ordinal association between the distance metrics and D-criteria. Mathematically, it is defined as:

\begin{equation}
\tau=\frac{2}{n(n-1)} \sum_{i<j} sgn\left(x_i-x_j\right) sgn\left(y_i-y_j\right),
\end{equation}

where $(x_{1},y_{1}),...,(x_{n},y_{n})$ are a set of samples of the variables.

$\tau$ is notable for its ability to measure the strength and direction of the relationship between two variables without requiring them to be on the same scale. Unlike parametric correlations like Pearson's, which assume linear relationships and normal distribution of data, Kendall's approach is based on the ranking of data points, assessing concordance and discordance in their relative ordering across two datasets. It focuses on rank rather than absolute values obviates the need for identical scales between datasets. Consequently, we can employ it to compare the results of the D-criteria and the distance metric without applying any normalization. We use the asymptotic method to compute Kendall’s tau, which provides an efficient and scalable approximation suitable for large datasets and handles ties effectively.

The process is as follows. For each meteor in the dataset, we first compute its similarity/closeness to every meteor shower based on predefined D-criteria and distance metrics (both for GEO and ORBIT vectors). These calculations yield two separate sets of rankings for every meteor: one set derived from the D-criteria and another from the distance metrics. Each set sorts all meteor showers from the most to the least similar to the meteor in question. Once we obtain these rankings, the $\tau$ is computed for each meteor, comparing the two sets of rankings to ascertain the degree of ordinal classification. For more information into the Kendall rank correlation coefficient applied here refer to \citet{Kendall1938, Fenwick1994, Hollander2013book}.

\subsubsection{Top-k Accuracy}

The heart of the present study centers on the evaluation of the classification accuracy of various D-criteria and distance metrics. To address this challenge, a unified methodology is imperative for the consistent application of statistical tests across all approaches under consideration. Despite the diversity in metrics and D-criteria, they converge on a singular objective: to quantify the association between a meteor and its corresponding meteor shower. As such, the Top-k accuracy is employed as a standardizing criterion to compare the overall accuracy among the various methods \citep{NIPS2009_ba2fd310}.

It quantifies the frequency with which the correct label class is included among the first k predicted labels. In the specific context, these labels denote the meteor showers associated with individual meteoroid impacts as classified by CAMS. For each meteor in the dataset, the similarities and distances are calculated in relation to all reference meteor showers. These values are subsequently sorted in ascending order to generate a ranked list. A successful classification in the Top-1 category occurs when the meteor shower with the minimum similarity or distance aligns with the meteor shower associated with the meteor in the CAM dataset. Similarly, a Top-5 success is recorded if the associated meteor shower is among the top five labels in the ranked list, and this extends analogously for other values of k.

In the present study, multiple tests encompassing Top-1, Top-5, and Top-10 accuracy are performed to evaluate the efficacy of D-criteria and distance metrics in associating a meteor with its originating meteor shower. This multi-tiered approach enables both a precise assessment of the top prediction (Top-1) and an evaluation of the model's capacity to identify a broader set of correct associations (Top-5 and Top-10). While one might assume that the Top-1 accuracy is paramount for meteor association, it is important to consider the significance of conducting Top-5 and Top-10 analyses. These extended evaluations yield insights into the efficacy of various ranking methodologies, going beyond mere concurrence with CAMS classifications. These analyses aid in contrasting the variability in rankings produced by different metrics. It is distinct when two metrics diverge at the Top-1 level yet converge within the Top-5, compared to a scenario where they diverge up to the Top-10.


\subsubsection{Kolmogorov-Smirnov Test}

The Kolmogorov-Smirnov test (K-S test) serves as a robust, non-parametric statistical method designed to assess the goodness-of-fit and equivalence of continuous, one-dimensional probability distributions. The test is particularly advantageous due to its distribution-free nature, making it applicable to datasets without the assumption of any specific distribution. The K-S test is employed in two primary contexts: the one-sample K-S test and the two-sample K-S test. The two-sample K-S aims to compare two empirical distributions and to determine if the two samples come from the same distribution. The K-S statistic $D$ is:

\begin{equation}
D = \sup_{x} | F_{1,n}(x) - F_{2,m}(x) |
\end{equation}

where $F_{1,n}(x)$ and $F_{2,m}(x)$ are the empirical distribution functions of the two samples of sizes $n$ and $m$, respectively. Here we follow the treatment explained in \citet{Hodges1958ArM}.

When applying the K-S test to Top-1 test results, interpreting the results sheds light on the comparative distributions of accuracy between classification methods. Failing to reject the null hypothesis $H_0$ indicates no statistically significant difference in accuracy distributions, but it does not affirm equivalence in method performance. Conversely, rejecting $H_0$ suggests a statistically significant difference, supporting the alternative hypothesis $H_1$ that the samples originate from distinct distributions. This outcome implies that $H_0$ does not adequately explain the observed data, with the decision to reject based on the significance level $\alpha$, set here at 0.05 for 95\% confidence.

\subsubsection{Top-1 Agreement}

Consider two classifiers tested on a dataset consisting of two equally sized classes. The first classifier might excel in identifying Class A but fail to recognize Class B, whereas the second classifier achieves the opposite, accurately identifying Class B while mistaking instances of Class A. Despite both classifiers reporting an overall accuracy of 50\%, their distinct performance on the individual classes reveals a divergent understanding and representation of the underlying patterns in the data. This example underscores the necessity of applying another test, as (1) Kendall's correlation assesses whether the order of rankings is similar between two sets of observations and (2) K-S is specifically focused on the shape of accuracy distributions rather than precise values.

For this reason, we calculate as well the percentage of Top-1 coincidence between distance metrics and D-vectors on an event-by-event basis, which provides a direct measure of agreement on the most preferred classification outcome, capturing the extent to which different approaches concur on the single best classification. This straightforward metric offers an immediate sense of the hit-and-miss between approaches. A heatmap is an optimal visualization tool for showcasing the pairwise agreement between classification metrics, using a rectangular matrix to highlight the magnitude of their coincidences.

\subsection{Differentiating the sporadic background}
\label{diff_sporadic}

The last part of our work deals with the effective discrimination of the sporadic background from meteor events that are associated with specific showers. We calculate the Top-1 accuracy values across the entire (non-filtered) database and construct the Receiver Operating Characteristic (ROC) curves for each D-criteria and distance metric, utilizing both the GEO and ORBIT vectors, using binary labels from CAMS (0: sporadic; 1: associated). The ROC curve represents the diagnostic ability of a binary classifier system as its discrimination threshold is varied. Using the ROC curve output, it is possible to quantify the optimal threshold that maximizes the classifier's performance with Youden's $J$ statistic \citep{Youden1950, Schisterman2005}:

\begin{equation}
J =  \frac{TP}{TP + FN}  + \frac{TN}{TN + FP} - 1,
\end{equation}

where TP represents the true positives, FN the false positives, TN the true negatives, and FP the false positives.

To synthesize the overall performance of each classification method in differentiating the sporadic background, we utilize the Matthews Correlation Coefficient, usually denoted by MCC or $\phi$ \citep{Matthews1975}. The $\phi$ offers a measure of the quality of binary classifications, encapsulating sensitivity, specificity, and the balance between them. It ranges from -1 (total disagreement between prediction and observation) to 1 (perfect prediction), with 0 denoting random guessing. The $\phi$ is defined as:

\begin{equation}
    \phi = \frac{(TP \cdot TN) - (FP \cdot FN)}{\sqrt{(TP + FP) \cdot (TP + FN) \cdot (TN + FP) \cdot (TN + FN)}}.
\end{equation}

\section{Results}
\label{results}

Similar to Section \ref{DataMethods}, where we detailed the database and methodology in distinct subsections, the results section is also organized into subsections for clarity and depth. Subsection \ref{population-analysis} examines the dataset, Subsection \ref{correlations} presents the rank correlation estimations, Subsection \ref{topk} reports on the accuracy results, Subsection \ref{equivalence} explores the equivalence between distance metrics and D-criteria, Subsection \ref{coincidence} offers the level of coincidence between approaches for the Top-1 tests, and finally Subsection \ref{perf_sporadic} provides optimal cut-offs and false positive rates.

\subsection{Population Analysis}
\label{population-analysis}

Within the extensive CAMS database, 24.6\% of its entries can be directly linked to a distinct meteor shower. In contrast, 75.4\% of the data points are categorized as sporadic events, implying they are part of the broader meteoroid background rather than specific meteor showers. After applying the filters mentioned in Section \ref{DataMethods}, the database reduces its number to account for 102,680 orbits.

The number of unique meteor shower classifications is somewhat constrained, amounting to 376 distinct categories. A total of 80\% of these classified meteor showers have been observed more than 10 times. A quarter of them, or 25\%, boasts over 100 individual recorded meteor events. An even smaller fraction, 5\%, can claim over 1000 meteor instances. Four of the meteor showers stand out due to their frequent documentation: the Perseids, Orionids, Geminids, and Southern Taurids (enumerated in descending order based on their observation frequency). Meteors belonging to these showers have been observed more than 10,000 times. 

Regarding the IAU meteor shower database, after filtering it reduces its number to 724, having 355 unique IDs shared with the CAMS database. Note that $\sim$30\% are duplicate entries, corresponding to distinct values for the same meteor shower estimated in different studies.

\begin{figure}
\centering
\includegraphics[width=0.7\textwidth]{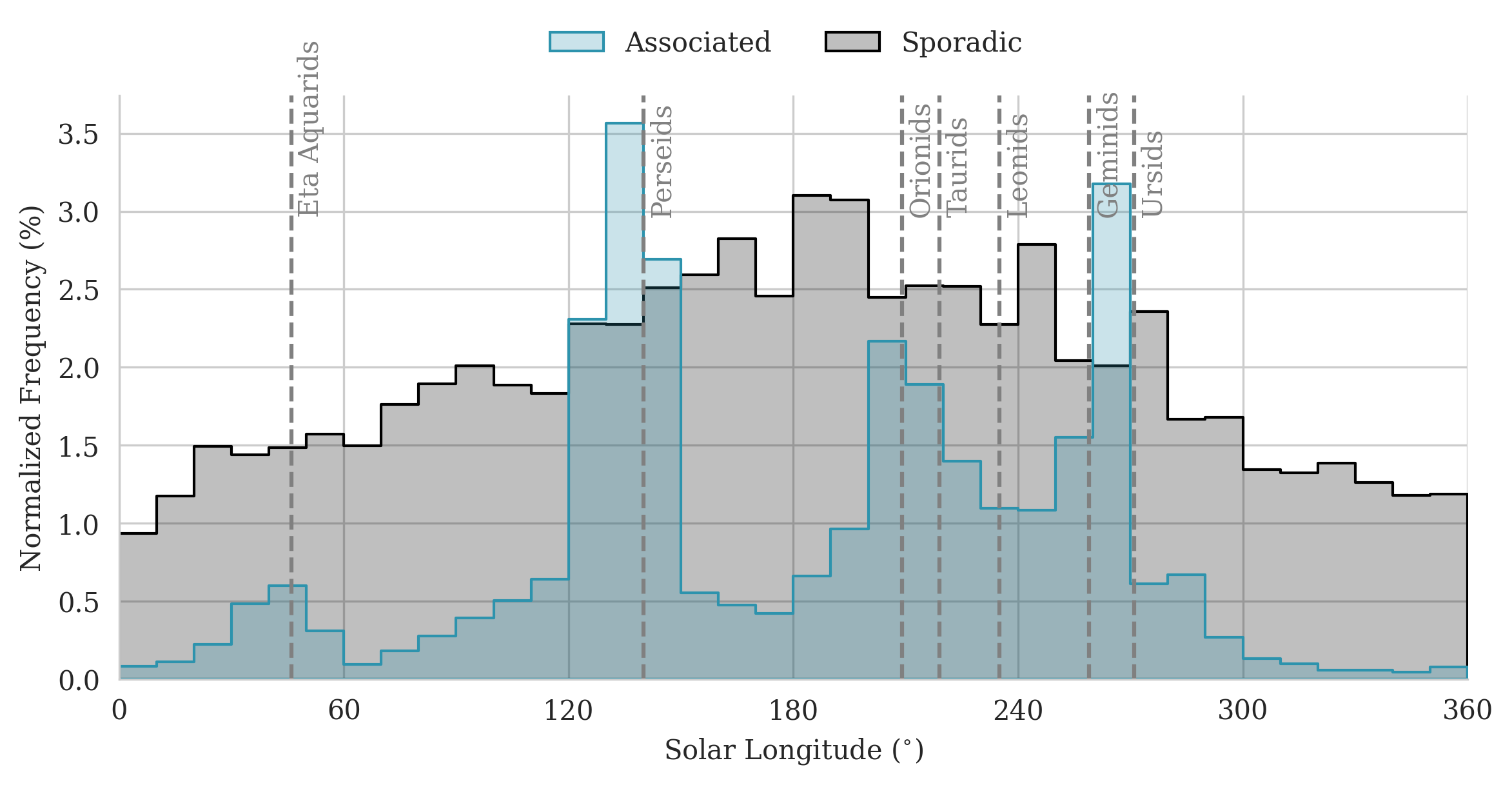}
\caption{Histogram of CAMS database as a function of the solar longitude. Sporadic and associated meteors are depicted.}
\label{fig:pop-freqlos}
\end{figure}

A key aspect of our analysis of associations is the parameter of solar longitude, that correlates meteor activity with Earth's specific orbital locations. Such a correlation is instrumental in discerning patterns and understanding recurring meteoritic phenomena. To visually represent this correlation, Figure \ref{fig:pop-freqlos} offers a histogram that plots impacting meteoroid classifications (sporadic or associated) based on solar longitude. The most active meteor showers are annotated. It can be observed an apparent concentration of the meteoroid background activity toward 180$^\circ$ of solar longitude.

\subsection{Degree of Rank Correlation}
\label{correlations}

For each of the showers listed in the IAU database, we compute the similarity/closeness between the shower and each meteor in the CAMS database using the D-criteria and all distance metric combinations. We then calculate the Kendall rank correlation between each D-criterion and each vector-metric combination. 

\begin{figure}
\centering
\includegraphics[width=1\textwidth]{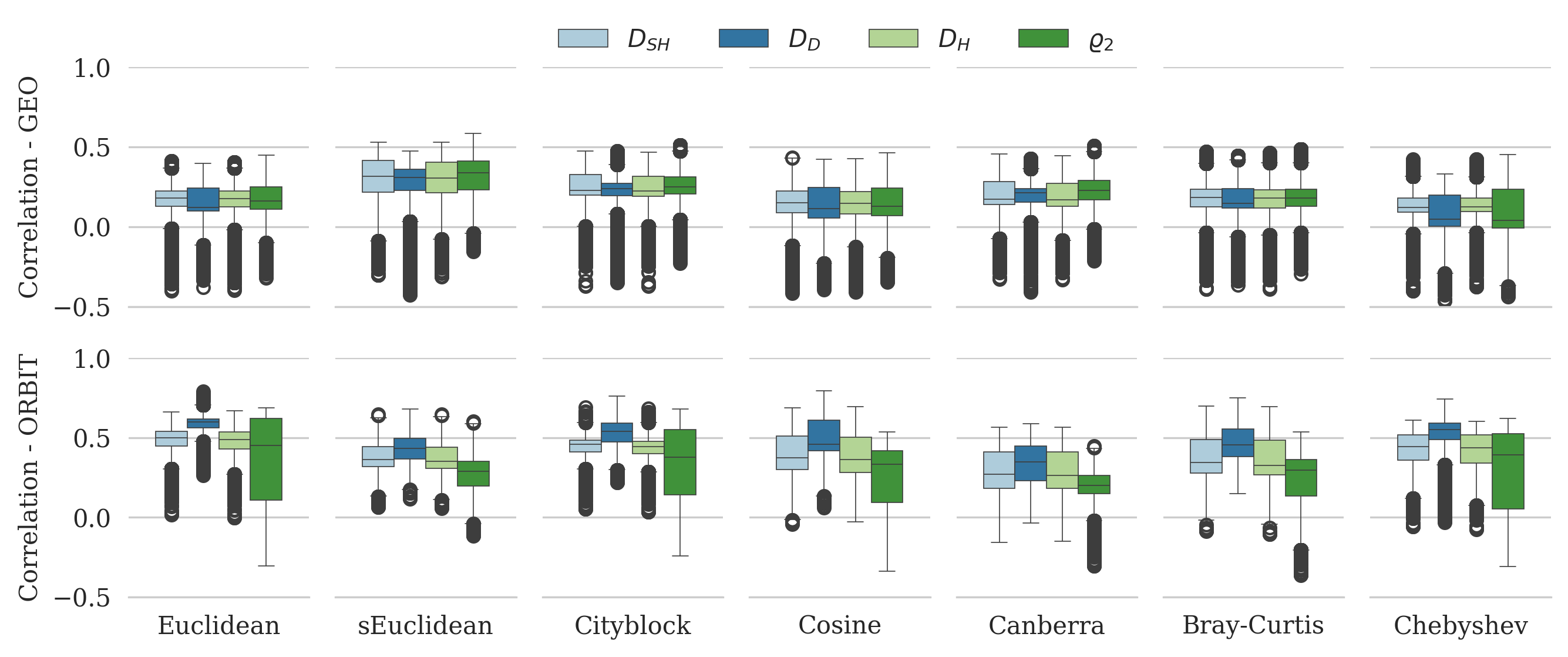}
\caption{Kendall rank correlation between D-criteria and metric distances for associated meteors in CAMS database. Each column corresponds to a unique vector (ORBIT or GEO). Each sample symbolizes the rank correlation between the similarity criteria and the distance metrics of each of the meteors from the CAMS database concerning the distinct meteor showers. Outlier values surpass 1.5 times the interquartile range of the median.}
\label{fig:corrs}
\end{figure}

Figure \ref{fig:corrs} displays the Kendall rank correlation between the evaluated D-criteria and distance metrics. Each column corresponds to a particular distance metric, and the plots are color-coded by D-vectors. The box plots encapsulate the quartile distribution of the samples, where each sample denotes the rank correlation between the D-criteria and distance metrics for a meteor with all meteor showers. The calculation is performed for each meteor against all meteor showers, a process executed separately for both the GEO and ORBIT vectors. Points lying outside the whiskers of the box plots are classified as outliers, positioned more than 1.5 times the interquartile range away from the median (Q2, depicted by the box's central line). A homogeneous dataset would result in a compact interquartile range, with the median equidistant from the box's extremes (Q1 and Q3), indicating symmetry. The span from the plot to each whisker indicates the data's variability or spread, suggesting a more concentrated distribution if the span is shorter and greater dispersion if it is extended.

The different figures reveal particular features in the Kendall rank correlation between D-criteria and distance metrics, as delineated by the employment of GEO and ORBIT vectors. The sEuclidean metric paired with the GEO vector consistently demonstrates the highest median correlation across all D-criteria, indicating a robust ordinal association. In contrast, the ORBIT vector presents a distinctive landscape. $D_D$ criterion, when evaluated with ORBIT vectors, achieves the highest correlation values. $\varrho_{2}$ criterion exhibits considerable variability in correlation, as evidenced by notably wide box plots for some distance metrics when using the ORBIT vectors. This behavior starkly contrasts with the other D-criteria, pointing to $\varrho_{2}$ unique response to the parameters captured by ORBIT vectors. While the GEO vector is characterized by a greater number of lower outliers, indicating instances of significantly divergent rankings, ORBIT vectors show fewer upper outliers. The results show a general tendency for the median correlation values to either be randomly centered or skewed across both vectors and all metrics. This variability suggests that no singular pattern of correlation prevails universally. Additionally, the maximum whisker extension observed with the Cosine distance metric, specifically when paired with the GEO vector and $D_D$ criterion, signals instances of high variability or dispersion in the degree of correlation.

\subsection{Accuracy of Best Choices}
\label{topk}

Using both the D-criteria and the employed meteor vectors and distance metrics, each meteor is juxtaposed with all showers, as detailed in Tables \ref{table:topk-dissim} and \ref{table:top1top5top10-vec}. This approach facilitates representing, in percentages, instances where the associated shower from the CAMS database in terms of distance and similarity aligns (Top-1) ranks among the five showers with the most minimal similarity and standardized Euclidean distance values (Top-5) or falls within the top ten showers (Top-10). 

\begin{table}
\centering
\caption{Top-k accuracies of D-criteria for associated meteors in CAMS database.}
\label{table:topk-dissim}
\begin{tabular}{ccccc}
  \hline
  \textbf{Top-k} & 
  \textbf{$\boldsymbol{D_{SH}}$} \textbf{(\%)} &
  \textbf{$\boldsymbol{D_{D}}$} \textbf{(\%)}&
  \textbf{$\boldsymbol{D_{H}}$} \textbf{(\%)}&
  \textbf{$\boldsymbol{\varrho_{2}}$ \textbf{(\%)}}
  \\ 
  \hline
    1 & \textbf{86.23} & 80.07 & 85.83 & 85.56 \\
    5 & 94.90 & 93.60 & 94.99 & \textbf{95.67} \\
    10 & 97.50 & 96.63 & 97.60 & \textbf{97.93} \\
  \hline
\end{tabular}
\end{table}

\begin{table}
\centering
\caption{Top-1, Top-5, and Top-10 accuracies of distance metrics for associated meteors in CAMS database.}
\label{table:top1top5top10-vec}
\begin{tabular}{cccccccc}
  \hline
  \textbf{} &
  \textbf{Euclidean} &
  \textbf{sEuclidean} &
  \textbf{Cityblock} &
  \textbf{Cosine} & 
  \textbf{Canberra} & 
  \textbf{Bray-Curtis} & 
  \textbf{Chebyshev} \\
  \hline
   & & & & Top-1 (\%) & & & \\
   \hline
    GEO & 84.68 & \textbf{87.06} & 84.79 & 84.63 & 85.89 & 84.96 & 83.41 \\ 
    ORBIT & 83.67 & 77.53 & 83.92 & 83.83 & 79.64 & \textbf{83.96} & 80.11 \\
  \hline
   & & & & Top-5 (\%) & & & \\
   \hline
    GEO & 92.99 & \textbf{95.03} & 93.72 & 92.95 & 94.35 & 93.89 & 90.94 \\ 
    ORBIT & \textbf{94.17} & 92.68 & 94.05 & 93.88 & 91.06 & 94.10 & 92.29 \\
  \hline
   & & & & Top-10 (\%) & & & \\
   \hline
    GEO & 95.56 & \textbf{97.06} & 96.16 & 95.52 & 96.66 & 96.23 & 94.05 \\ 
    ORBIT & 96.37 & 96.25 & 96.57 & 96.19 & 95.08 & \textbf{96.61} & 94.45 \\
    \hline
\end{tabular}
\end{table}

The optimal D-criterion for achieving Top-1 accuracy is $D_{SH}$ (86.23\%), whereas $\varrho_{2}$ excels in both Top-5 (95.67\%) and Top-10 (97.93\%) categories and have a good accuracy in Top-1 (85.56\%). Conversely, $D_{D}$ ranks as the least effective across all evaluations, markedly lagging behind the others, which exhibit comparable performances. The sEuclidean when combined with the GEO vector demonstrates superior performance (87.06\%) over the other D-criteria in achieving Top-1 accuracy and overly the rest of the distance metrics in all Top-k tests. When paired with the ORBIT vector, the Bray-Curtis metric delivers the highest overall accuracy (including the $D_{D}$ criterion in all tests), except for Top-5 accuracy, where the Euclidean metric slightly outperforms it. Across the distance metrics evaluated, the GEO vector is found to yield better outcomes than the ORBIT vector. The Chebyshev metric exhibits the worst results with the GEO vector, while the sEuclidean and Canberra present the lower performances for the ORBIT vector. Table \ref{tab:accuracy_mean} shows the mean accuracies. The distance metrics combined with the GEO vectors offer the best overall accuracy for Top-1, while the D-criteria outstrip in Top-5 and Top-10.

\begin{table}[ht]
\centering
\caption{Mean accuracies and standard deviations for Top-k tests across the D-criteria and distance metrics with GEO and ORBIT vectors.}
\label{tab:accuracy_mean}
\begin{tabular}{lcccc}
\hline
\textbf{Test} & \textbf{All (\%)} & \textbf{D-criteria (\%)} & \textbf{GEO (\%)} & \textbf{ORBIT (\%)} \\
\hline
Top-1 & \(83.7 \pm 2.5\) & \(84.2 \pm 2.5\) & \(\boldsymbol{85.1\pm1.1}\) & \(81.8 \pm 2.5\) \\
Top-5  & \(93.6 \pm 1.3\) & \(\boldsymbol{94.8 \pm 0.7}\) & \(93.4 \pm 1.2\) & \(93.2 \pm 1.1\) \\
Top-10 & \(96.2 \pm 1.0\) & \(\boldsymbol{97.4 \pm 0.5}\) & \(95.9 \pm 0.9\) & \(95.9 \pm 0.8\) \\
\hline
\end{tabular}
\end{table}

\begin{figure}
\centering
\includegraphics[width=0.7\textwidth]{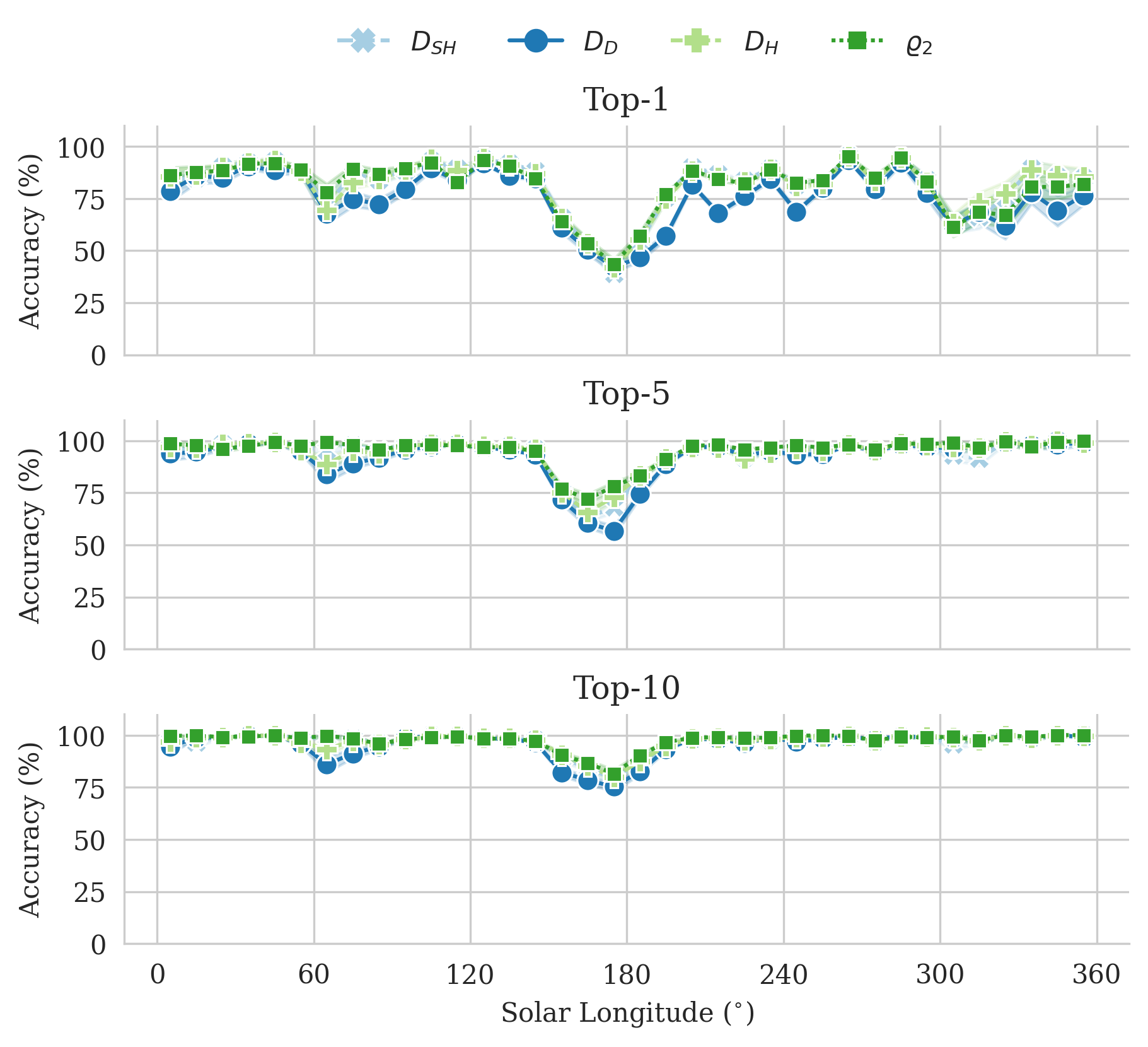}
\caption{Top-k accuracies along solar longitude of the D-criteria for associated meteors in CAMS database.}
\label{fig:topk-diss}
\end{figure}

\begin{figure}
\centering
\includegraphics[width=1\textwidth]{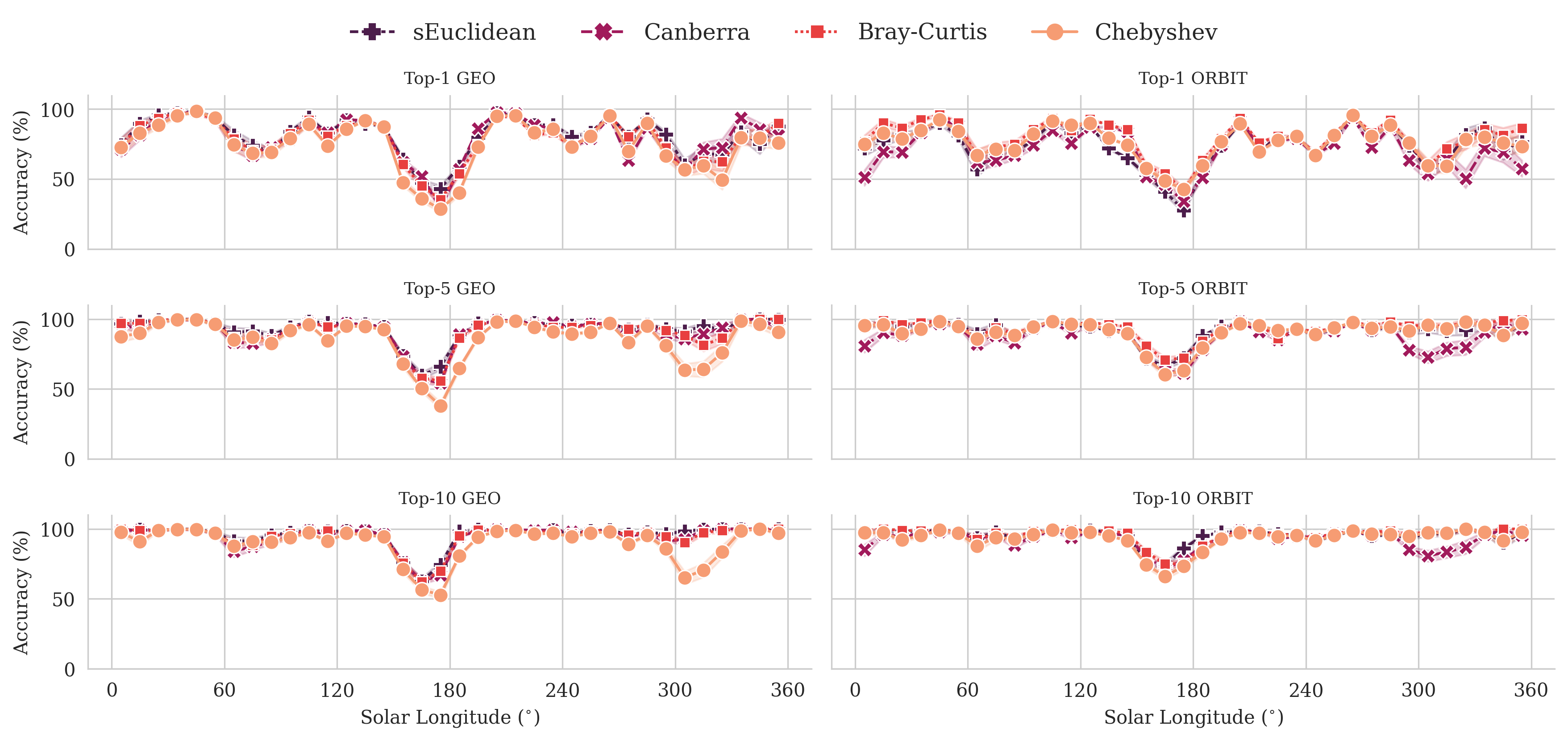}
\caption{Top-k accuracies along solar longitude of different distance metrics for associated meteors in CAMS database.}
\label{fig:topk-orbit}
\end{figure}

Figure \ref{fig:topk-diss} illustrates the variation in Top-k test accuracy as a function of solar longitude across different D-criteria. Similarly, Figure \ref{fig:topk-orbit} displays the Top-k results for the sEuclidean, Canberra, Bray-Curtis, and Chebyshev distance metrics. Across all evaluations, the results are of the same order of magnitude. A distinct pattern emerges: for Top-1, the accuracy variation is irregular, whereas, for Top-10, it tends towards uniformity, except for a notable decrease (up to 50\%) around 180$^\circ$ of solar longitude. Visually, the lower performance of $D_D$ is prominent, and $D_{SH}$ and $\varrho_{2}$ excel, especially at solar longitudes between 170$^\circ$ and 220$^\circ$, as well as around 70$^\circ$ in Top-5 and Top-10, and 350$^\circ$ in Top-1 (with a sudden increase of the accuracy of $D_H$). Conversely, the performances of the distance metrics generally follow the same trend, albeit less uniformly in the Top-10 distribution. Besides the common peak at 180$^\circ$, it is observed that they struggle to associate meteors at around 310$^\circ$, where Chebyshev (with GEO vector) and Canberra (with ORBIT vector) exhibit remarkably lower performances.

\begin{figure}
\centering
\includegraphics[width=0.7\textwidth]{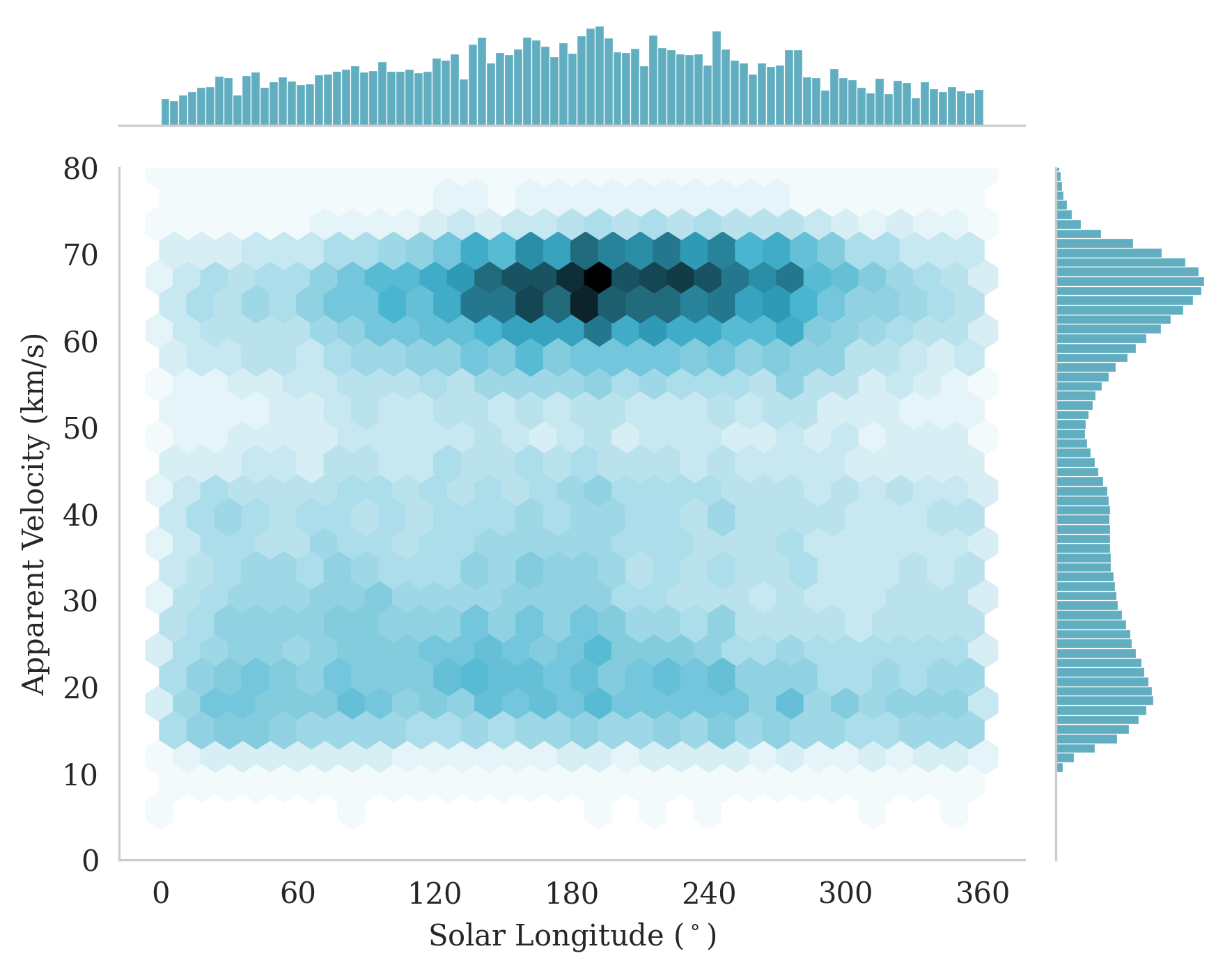}
\caption{2D-histogram of sporadic meteor apparent velocities and solar longitudes at impact in the CAMS database. Darker colors denote higher density.}
\label{fig:vel_sol_spo}
\end{figure}

The trend of minimum accuracy in meteor association is pinpointed at 180$^\circ$ solar longitude, aligning with an apparent increase in the meteoroid background activity, as depicted in Fig. \ref{fig:pop-freqlos}. This time frame also bridges the Perseids and Orionids, meteor showers renowned for their high activity and velocities above 60 km/s, expecting, in consequence, a diffuseness of their parameters. Instrumental constraints correlate meteoroid velocity with measurement inaccuracies \citep{Hajdukova2020PSS}. As a result, high-velocity meteoroids are more challenging to accurately characterize. This is depicted in Figure \ref{fig:vel_sol_spo}, showcasing a concentration of high apparent velocities within this specific solar longitude range. It is conceivable that these meteoroids were once part of such swarms but have lost their orbital affinity due to temporal decoherence, making many of them challenging to distinguish. Furthermore, the increased activity during these periods, characterized by similar velocities, may have influenced the association process conducted by CAMS.

\subsection{Statistical Equivalence}
\label{equivalence}

\begin{figure}
\centering
\includegraphics[width=0.7\textwidth]{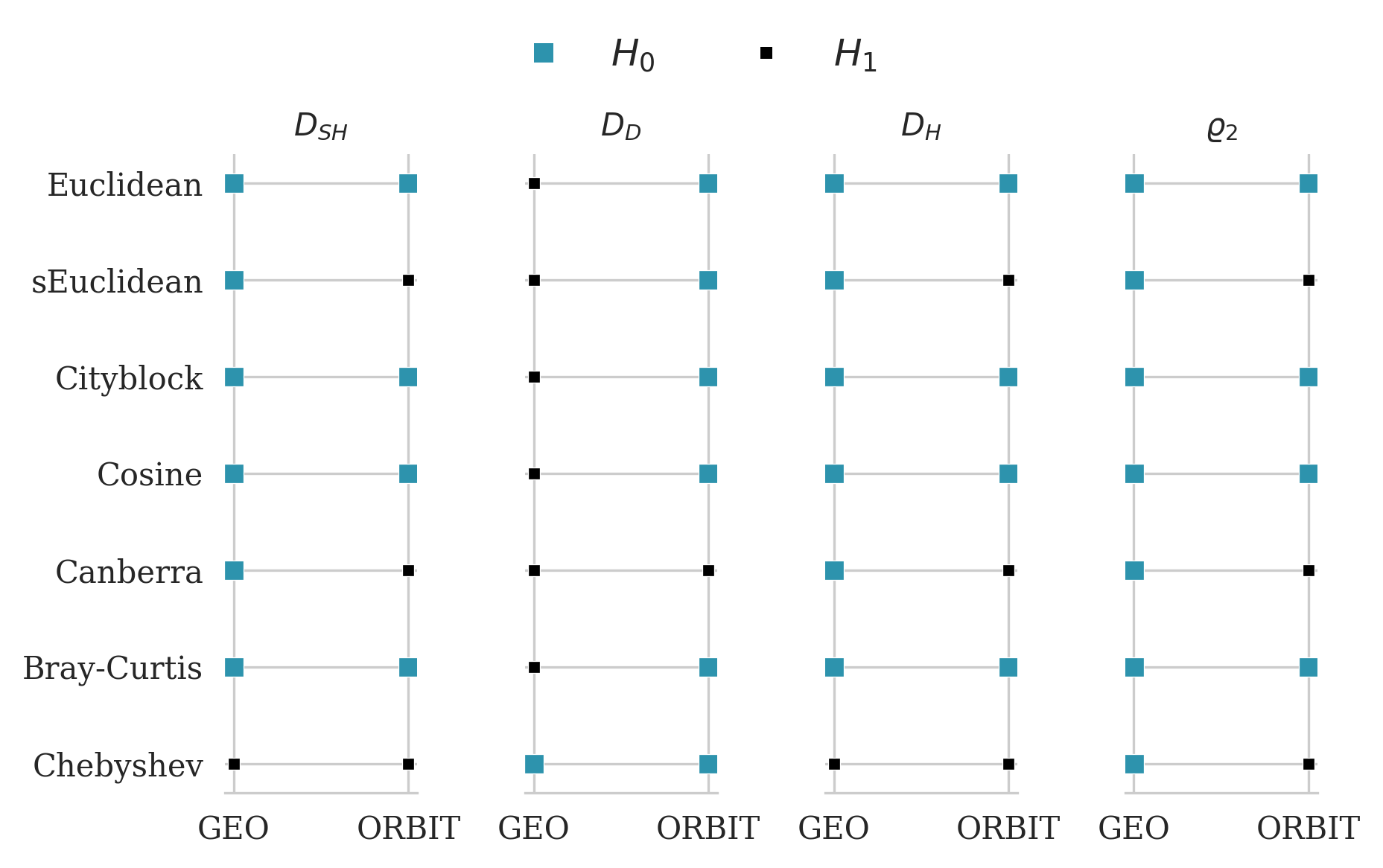}
\caption{K-S test comparing Top-1 accuracies of distance metrics and D-criteria with a 95\% level of confidence for associated meteors in CAMS database. $H_0$ indicates no statistically significant difference between distributions, while $H_1$ indicates a significant difference between the compared distributions.}
\label{fig:2ks-test}
\end{figure}

Figure \ref{fig:2ks-test} displays classification outcomes labeled as $H_0$ or $H_1$, corresponding to the hypothesis tested for each data comparison for the Top-1 accuracy results from the metric distance and the D-criteria. Labels are determined based on p-values: instances where the p-value is less than 0.05 are marked as $H_1$, indicating the rejection of the null hypothesis ($H_0$) in favor of the alternative ($H_1$), suggesting a statistically significant difference between the compared distributions. Conversely, instances with a p-value greater than or equal to 0.05 retain the $H_0$ label, indicating insufficient evidence to reject the null hypothesis, thus suggesting no statistically significant difference between distributions under examination.

The figure reveals a distinct pattern in the distribution of hypothesis testing results, particularly when evaluating the $D_D$ criterion with the GEO vector. Contrary to the other D-criteria, which generally do not reject the null hypothesis $H_0$ when paired with the GEO vector, $D_D$ stands out by predominantly rejecting $H_0$ (indicated by $H_1$), suggesting differences in distributions. This trend is reversed for the ORBIT vector, where $D_D$ results in non-rejection of $H_0$, except for one the Canberra metric. This behavior is markedly different from the other criteria tested with the ORBIT vector, which rejects $H_0$ for the same three metrics (sEuclidean, Canberra, and Chebyshev). $\varrho_{2}$ appears the most likely compatible with both vectors at the same time.

The consistent failure to reject $H_0$ with GEO vector for all distance metrics under the $\varrho_{2}$ criterion does not confirm the distributions being identical but rather indicates the test lacked sufficient evidence to demonstrate statistical differences. This outcome positions $\varrho_{2}$ as the D-criterion that is most plausibly comparable to the distance metrics in terms of meteor association when using the GEO vector. Also, the $\varrho_{2}$ metric exhibits the highest probability of being compatible with both vectors simultaneously.

\subsection{Event-by-Event Agreement}
\label{coincidence}

\begin{figure}
\centering
\includegraphics[width=0.9\textwidth]{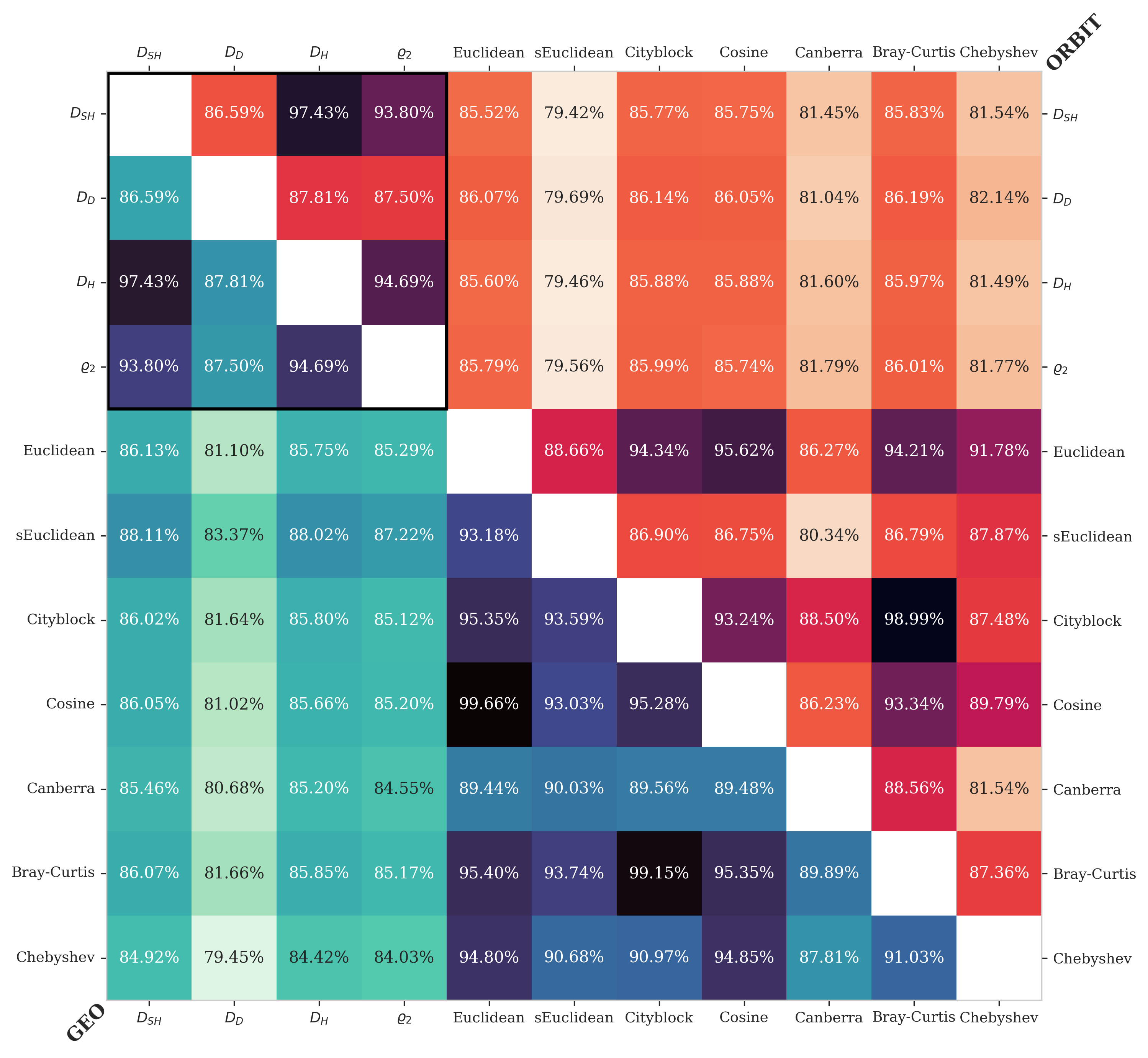}
\caption{Heatmap of cross-coincidence between D-vectors and distance metrics using GEO (lower triangle, blue colormap) and ORBIT (upper triangle, red colormap) vectors of Top-1 accuracies for associated meteors in CAMS database. D-vector's own cross-coincidences are highlighted within a black rectangle in the top left corner.}
\label{fig:coinc}
\end{figure}

The heatmap on Figure \ref{fig:coinc} visualizes the agreement level between various D-vectors and distance metrics, showcasing their comparative analysis for the Top-1 results across the two meteor vectors. Each cell represents the percentage of Top-1 coincidence between pairs, with GEO-related comparisons highlighted in shades of blue for intuitive analysis, and ORBIT-related comparisons in shades of red, enabling a clear distinction between the two meteor vectors used. The diagonal, intentionally left blank, separates GEO and ORBIT results for a dual analysis within a single visual representation. The cross-accuracies of D-criteria, independent of GEO or ORBIT vectors, are outlined with a black frame in the figure's top left corner.

Analyzing the heatmap reveals that the D-vector $D_{SH}$ has a strong event-by-event alignment with $D_{H}$ for Top-1 (97.43\%), indicating these criteria frequently concur on their top classifications. This is closely followed by $\varrho_{2}$ and $D_{H}$ (94.69\%). Within the GEO vector, Euclidean and Cosine (99.66\%), along with Cityblock and Bray-Curtis (99.15\%), show the highest levels of coincidence in Top-1. The sEuclidean metric generally shows good agreement for the GEO vector across various metrics and D-criteria ($\sim$88\%), except when paired with $D_D$ (83.37\%). For the ORBIT vector, Cityblock and Bray-Curtis (98.99\%), as well as Euclidean and Cosine (95.62\%), exhibit the highest values. There is better alignment between $D_D$ and the ORBIT vector (reaching $\sim$86\% with various distance metrics) than seen with the GEO vector.

The consistency observed in the heatmap resonates with the findings from Kendall's correlation and the K-S test. These statistical measures support the identified patterns of agreement and discrepancy among the classifiers, providing robustness to the analysis and confirming the reliability of these patterns.

\subsection{Thresholds and Confusion Matrices}
\label{perf_sporadic}

Table \ref{table:metrics-vec} presents the evaluation of D-criteria and distance metrics within the CAMS database, considering both sporadic and associated meteor events, where optimal thresholds and the effectiveness of different methods are encapsulated. The standout performer among D-criteria is $D_{SH}$, distinctly outshining others with a $\phi$ of 0.6400. Conversely, $D_{D}$ emerges as the least effective.

When using the GEO vector, the sEuclidean metric takes precedence, exhibiting the highest overall accuracy and a $\phi$ value of 0.6464, closely followed by Cityblock and Bray-Curtis metrics. The scenario shifts when transitioning to the ORBIT vector, where Cityblock edges out as the frontrunner, albeit with Bray-Curtis and Euclidean not far behind, suggesting a competitive field with closely matched performances. The sEuclidean metric with ORBIT vector does not mirror its GEO vector success, hinting at vector-specific behavior that influences metric efficacy.

Cityblock, while not outperforming other distance metrics in replicating CAMS' associated Top-1 classifications, excelled in more effectively distinguishing the sporadic background on average. Except for $D_{SH}$, all distance metrics applied to the GEO vector--aside from Cosine--surpass the rest of the D-criteria in terms of the $\phi$. Interestingly, despite generally lower performance with the ORBIT vector, several distance metrics still exceed some D-criteria performances. Cityblock, in particular, scores relatively close to achieving the superior results of $D_{SH}$ and sEuclidean.

Additionally, the observed thresholds for traditional D-criteria ($D_{SH}$, $D_{D}$, and $D_{H}$) align perfectly with values documented in the scientific literature, reinforcing the validity of our findings.

\begin{table}[h]
\centering
\caption{Threshold, accuracies, and Matthews correlation coefficients for different D-criteria and distance metrics in the CAMS database taking into account the sporadic and associated events.}
\label{table:metrics-vec}
\begin{tabular}{lccccccc}
  \hline
  \textbf{Method} & \textbf{Cut-off} & \textbf{TP (\%)} & \textbf{FN (\%)} & \textbf{FP (\%)} & \textbf{TN (\%)} & \textbf{Acc. (\%)} & $\boldsymbol{\phi}$ \\
  \hline
  \multicolumn{8}{c}{D-criteria} \\
  \hline
$D_{SH}$  &  0.170  &  59.51  &  8.62  &  7.21  &  24.66  &  84.17  &  \textbf{0.6400} \\
$D_{D}$  &  0.083  &  58.34  &  9.78  &  8.35  &  23.52  &  81.87  &  0.5877 \\
$D_H$  &  0.176  &  58.81  &  9.32  &  7.76  &  24.12  &  82.92  &  0.6122 \\
$\varrho_2$  &  0.174  &  59.56  &  8.56  &  7.97  &  23.91  &  83.47  &  0.6213 \\
  \hline
  \multicolumn{8}{c}{Distance metric with the GEO vector} \\
  \hline
Euclidean  &  0.138  &  58.75  &  9.37  &  7.13  &  24.75  &  83.50  &  0.6279 \\
sEuclidean  &  0.370  &  59.25  &  8.88  &  6.78  &  25.10  &  84.34  &  \textbf{0.6464} \\
Cityblock  &  0.252  &  59.45  &  8.67  &  7.20  &  24.67  &  84.13  &  0.6393 \\
Cosine  &  0.004  &  57.85  &  10.27  &  6.97  &  24.91  &  82.76  &  0.6154 \\
Canberra  &  0.380  &  59.98  &  8.15  &  8.22  &  23.65  &  83.63  &  0.6228 \\
Bray-Curtis  &  0.037  &  59.80  &  8.33  &  7.48  &  24.40  &  84.19  &  0.6387 \\
Chebyshev  &  0.089  &  61.01  &  7.11  &  8.83  &  23.05  &  84.06  &  0.6282 \\
  \hline
  \multicolumn{8}{c}{Distance metric with the ORBIT vector} \\
  \hline
Euclidean  &  0.086  &  60.23  &  7.89  &  8.01  &  23.86  &  84.10  &  0.6335 \\
sEuclidean  &  0.542  &  59.80  &  8.32  &  9.71  &  22.17  &  81.97  &  0.5803 \\
Cityblock  &  0.173  &  59.63  &  8.50  &  7.46  &  24.42  &  84.04  &  \textbf{0.6359} \\
Cosine  &  0.002  &  57.47  &  10.65  &  6.94  &  24.94  &  82.41  &  0.6093 \\
Canberra  &  0.431  &  55.69  &  12.43  &  8.13  &  23.74  &  79.44  &  0.5454 \\
Bray-Curtis  &  0.031  &  59.31  &  8.81  &  7.29  &  24.58  &  83.90  &  0.6343 \\
Chebyshev  &  0.061  &  59.55  &  8.57  &  8.54  &  23.34  &  82.89  &  0.6062 \\
  \hline
\end{tabular}
\end{table}

As an additional note to our findings, it is noteworthy that upon incorporating the complete list of meteor showers—not limited to those used within the CAMS database—an average of 27\% of the meteor classified (Top-1) by all D-criteria and distance metrics would align better with newly recognized meteor showers. In future efforts, we aim to do a comparative analysis by testing other databases such as GMN \citep{Vida2021MNRAS} and EDMOND \citep{Kornos2014pim3}.

\section{Conclusions}
\label{conclusions}

This study undertook a statistical evaluation of four orbital similarity criteria (or D-criteria) within a five-dimensional parameter space to probe the dynamical associations within meteor data. Utilizing the extensive data compiled by the CAMS network, we have not only relied on D-criteria ($D_{SH}$, $D_D$, $D_H$, and $\varrho_2$) but also ventured into distance metrics commonly applied in Machine Learning (Euclidean, sEuclidean, Cityblock, Cosine, Canberra, Bray-Curtis, and Chebyshev), investigated across two distinctive meteor vectors. One vector termed ORBIT, based on heliocentric orbital elements, is essentially shared with the D-criteria, and the other one, GEO, based on geocentric observational parameters, was proposed by \citet{Sugar2017MPS}. Our methodology hinged on the Kendall rank correlation coefficient and Top-k accuracy tests to assess the correlation and performance of these criteria and metrics. We also applied the Kolmogorov-Smirnov test and computed the level of coincidence of individual Top-1 results for discerning the statistical equivalence of the different approaches. Finally, we calculated the optimal thresholds and evaluated their performances in distinguishing the sporadic background from the meteor showers.

Our key findings can be summarized as follows:
\begin{itemize}
    \item The sEuclidean metric paired with the GEO vector demonstrates superior performances than the D-criteria and the other distance metrics in achieving Top-1 accuracy (87.06\%).
    
    \item Regarding the D-criteria, the $D_{SH}$ criterion holds the upper hand in achieving Top-1 accuracy (86.23\%), while $\varrho_2$ maintains dominance in both the Top-5 (95.67\%) and Top-10 (97.93\%) categories (surpassed by $D_{SH}$ in Top-1 accuracy by 0.67\%).
    
    \item The Bray-Curtis metric, allied with the ORBIT vector, demonstrated a consistent edge over other distance metrics, outperforming the $D_{D}$ criterion across all Top-k tests (83.96\%, 94.10\%, and 96.61\%, in increasing order of k) and only slightly beaten by the Euclidean metric in Top-5 accuracy by a negligible difference (0.07\%).
    
    \item $D_D$ exhibits an opposite trend to the other D-criteria when evaluating its equivalence against distance metrics with the GEO vector.
    
    \item Among the D-criteria, $\varrho_2$ appears as the most likely similar to the distance metrics with the GEO vector, being also the most compatible with both GEO and ORBIT vectors at the same time.
    
    \item In general terms, the D-criteria and the metric distances provide similar accuracies in Top-k tests (83.7$\pm$2.5\%, 93.6$\pm$1.3\%, 96.2$\pm$1.0\%, in ascending order of k), with the $D_D$ and the metric Chebyshev performing worse.
    
    \item The mean highest accuracies are achieved with the GEO (85.1$\pm$1.1\%, 93.4$\pm$1.2\%, and 95.9$\pm$0.9\%) rather than the ORBIT vector (81.8$\pm$2.5\%, 93.2$\pm$1.1\%, and 95.9$\pm$0.8\%) within the Top-1, Top-5, and Top-10 tests, and surpassing also the D-criteria in Top-1 (84.4$\pm$2.5\%, 94.8$\pm$0.7\%, and 97.4$\pm$0.5\%). This suggests that geocentric parameters provide a more robust basis than orbital elements for meteor dynamical association.
    
    \item We observed moderate solar longitude-dependent deviations and a common significant decrease in accuracy around 180$^\circ$ of solar longitude. We tentatively linked these features to heightened meteoroid background activity and the interface with two of the most active, high-velocity meteor showers: the Perseids and the Orionids.
    
    \item The sEuclidean metric achieves the highest overall accuracy (84.34\%) and $\phi$ of 0.6464 for GEO vector applications, excelling in distinguishing the sporadic background. It is closely followed by Cityblock ($\phi$=0.6393 and 84.13\% accuracy) and Bray-Curtis metrics ($\phi$=0.6387 and 84.19\% accuracy).
    
    \item Among D-criteria, $D_{SH}$ distinguishes itself with a $\phi$ of 0.6400, translating to an 84.17\% accuracy rate in separating the background, while $D_{D}$ emerges as the least effective, with a $\phi$ of 0.5877 and an accuracy of 81.87\%.
    
    \item In the ORBIT vector context, Cityblock takes the lead ($\phi$=0.6359 and 84.04\% accuracy), closely challenged by Bray-Curtis ($\phi$=0.6343 and 83.90\% accuracy) and Euclidean ($\phi$=0.6335 and 84.10\% accuracy).
    
    \item Excluding Cosine, all distance metrics associated with the GEO vector surpass the D-criteria in $\phi$ when differentiating the meteoroid background.
    
    \item Despite the ORBIT vector's generally lower performance, various distance metrics still exceed certain D-criteria in effectiveness.
    
    \item Optimal cut-offs for all D-criteria and distance metrics are provided, founded on the CAMS database classification.
    
    \item Based on these approaches, $\sim$27\% of associated meteors in CAMS would align with showers identified after the database's release.
    
    \item Future research will concentrate on studying effectiveness, equivalences, and thresholds for a synthetic impacting population, exploring the performance and specific attributes of the methods for each individual meteor shower.
\end{itemize}

The work culminates in the significant revelation that Machine Learning distance metrics can rival or even outperform the specifically tailored orbital similarity criteria for meteor dynamical association. This opens up new pathways for the use of computational techniques in the field of meteor science, offering an opportunity to refine our approaches to classifying meteor showers and sporadic meteors alike.




\subsection*{Acknowledgements}
\label{acknowledgements}

EP-A and JMS-L have carried out this work in the framework of the project Fundación Seneca (22069/PI/22), Spain. EP-A thanks funding from the European Research Council (ERC) under the European Union’s Horizon 2020 Research and Innovation Programme (grant agreement No. 865657) for the project “Quantum Chemistry on Interstellar Grains” (QUANTUMGRAIN). JMS-L thanks funding from the Ministerio de Ciencia e Innovación (Spain) (grants PID2020-112754GB-I00 and PID2021-128062NB-I00). This work was supported by the LUMIO project funded by the Agenzia Spaziale Italiana. We acknowledge Andrés Suárez-García for his initial involvement and contributions to this work.



\bibliographystyle{jasr-model5-names}
\biboptions{authoryear}
\bibliography{refs}

\end{document}